\documentclass[%
reprint,
superscriptaddress,
showpacs,preprintnumbers,
amsmath,amssymb,
prb,
]{revtex4-1}

\usepackage{graphicx}
\usepackage{dcolumn}
\usepackage{bm}
\usepackage{color}
\usepackage[colorinlistoftodos]{todonotes} 
\usepackage[export]{adjustbox}

\begin{document}

\newcommand{\mucm}{$\mathrm{\mu\Omega\, cm}$}
\newcommand{\RH}{$R_H$}


\title{Extreme Magnetoresistance in the Topologically Trivial Lanthanum Monopnictide: LaAs}

\author{H.-Y. Yang}
\affiliation{Department of Physics, Boston College, Chestnut Hill, MA 02467, USA}
\author{T. Nummy}
\affiliation{Department of Physics, University of Colorado, Boulder, CO 80309, USA}
\author{H. Li}
\affiliation{Department of Physics, University of Colorado, Boulder, CO 80309, USA}
\author{S. Jaszewski}
\affiliation{Department of Physics, Boston College, Chestnut Hill, MA 02467, USA}
\author{M. Abramchuk}
\affiliation{Department of Physics, Boston College, Chestnut Hill, MA 02467, USA}
\author{D. S. Dessau}
\affiliation{Department of Physics, University of Colorado, Boulder, CO 80309, USA}
\author{Fazel Tafti}
\email{fazel.tafti@bc.edu}
\affiliation{Department of Physics, Boston College, Chestnut Hill, MA 02467, USA}

\date{\today}

\begin{abstract}
The family of binary Lanthanum monopnictides, LaBi and LaSb, have attracted a great deal of attention as they display an unusual “extreme” magnetoresistance (XMR) that is not well understood.
Two classes of explanations have been raised for this:
the presence of non-trivial topology, and the compensation between electron and hole densities.
%
Here, by synthesizing a new member of the family, LaAs, and performing transport measurements, Angle Resolved Photoemission Spectroscopy (ARPES), and Density Functional Theory (DFT) calculations, we show that
%
%
(a) LaAs retains all qualitative features characteristic of the XMR effect but with a siginificant reduction in magnitude compared to LaSb and LaBi,
%
(b) the absence of a band inversion or a Dirac cone in LaAs indicates that topology is insignificant to XMR,
%
%
(c) the equal number of electron and hole carriers indicates that compensation is necessary for XMR but does not explain its magnitude, and
(d) the ratio of electron and hole mobilities is much different in LaAs compared to LaSb and LaBi.
We argue that the compensation is responsible for the XMR profile and the mobility mismatch constrains the magnitude of XMR.

\end{abstract}

\maketitle



\section{\label{Introduction}Introduction}

Semimetals are characterized by small and often compensated electron and hole carrier densities ($n_e/n_h \approx 1$) \cite{issi_low_1979}.
In elemental semimetals, such as bismuth, compensation between high mobility electron and hole carriers reduces the Hall field and produces a large magnetoresistance $\mathrm{MR}(\%)=100\times \left[ \rho(H)-\rho(0) \right] / \rho(0)$ \cite{mcclure_linear_1968,yang_large_1999,kopelevich_reentrant_2003,du_metal-insulator-like_2005,fauque_electronic_2009}.
A reduced Hall field fails to counteract the Lorentz force that bends the trajectory of charge carriers in a magnetic field, therefore results in a large MR \cite{issi_low_1979}.
An extremely large and non-saturating magnetoresistance with magnitude $\sim 10^{4-6}\%$ has been recently reported in several topological semimetals (TSMs) including WTe$_2$ \cite{ali_large_2014, zhao_anisotropic_2015}, Cd$_3$As$_2$ \cite{liang_ultrahigh_2015}, PtSn$_4$ \cite{mun_magnetic_2012, wu_dirac_2016}, NbSb$_2$ \cite{wang_anisotropic_2014}, NbAs \cite{ghimire_magnetotransport_2015}, NbAs$_2$ \cite{yuan_large_2016}, NbP \cite{shekhar_extremely_2015}, TaSb$_2$ \cite{li_resistivity_2016}, TaAs \cite{zhang_electron_2017}, TaAs$_2$ \cite{yuan_large_2016,luo_anomalous_2016}, and TaP \cite{zhang_large_2015}.
TSMs are extensions of topological insulators (TIs) where degenerate crossings between several bulk bands are protected by a fundamental symmetry of the material \cite{burkov_topological_2016}.
The $\rho(T)$ profile of the extreme magnetoresistance (XMR) in TSMs looks similar to the $\rho(T)$ profile of TIs where by decreasing temperature, resistivity shows an \emph{upturn} followed by a \emph{plateau} \cite{ren_large_2010,kim_surface_2013}.
In TIs, the upturn is assigned to a metal-insulator transition and the plateau is assigned to topological surface states.
The similarity between the XMR profile and the TI profile caused confusion and opened a debate over the possibility of XMR profile being rooted in the topological properties of TSMs \cite{ali_large_2014,liang_ultrahigh_2015,shekhar_extremely_2015,zhang_large_2015}.
Here, we try to settle this debate by making a new material which is topologically trivial but shows the typical XMR profile.
Lanthanum monopnictides (LaSb  and LaBi) \cite{tafti_resistivity_2016,tafti_temperaturefield_2016, sun_large_2016, kumar_observation_2016} have attracted special attention among XMR semimetals due to their simple cubic structure \cite{tafti_resistivity_2016}.
It has been shown that both LaSb and LaBi are compensated \cite{sun_large_2016,zeng_compensated_2016} but Dirac cones have also been observed clearly in LaBi \cite{niu_presence_2016,wu_asymmetric_2016,nayak_multiple_2017,lou_evidence_2017} and less clearly in LaSb\cite{niu_presence_2016,zeng_compensated_2016,oinuma_three-dimensional_2017} by ARPES.
Therefore, it is challenging to disentangle compensation from topology in relation to XMR by focusing on LaSb and LaBi.
The disagreement on the presence of Dirac cones in LaSb from ARPES results suggests a topological/non-topological transition within the lanthanum monopnictide family by decreasing the pnictogen size.
This observation motivated us to grow single crystals of LaAs with the hope of observing XMR in the absence of topological features.
Our detailed transport measurements, DFT calculations, and ARPES experiments reveal two important findings:
First, LaAs lacks a Dirac cone unambiguously, yet it exhibits the typical XMR transport profile.
Therefore, XMR is independent of topological character.
Second, LaAs is as compensated as LaSb and LaBi, but the magnitude of XMR in LaAs is orders of magnitude smaller.
Therefore, compensation is necessary to explain the presence of XMR but not sufficient to determine its magnitude.
Our results suggest that the relative mobilities of electrons and holes determine the magnitude of XMR in compensated semimetals.
Previous reports on the synthesis of LaAs are limited to polycrystalline samples \cite{hulliger_superconductivity_1977,shirotani_pressure-induced_2003}, thin films \cite{krivoy_growth_2012}, or mixed phases of LaAs$_2$/LaAs \cite{murray_halide_1970}.
This is the first report on the growth and characterization of pure LaAs single crystals.


\section{\label{Experiments}Methods}

 LaAs crystals were grown using a flux method as described in the Supplemental Material \cite{suppmatt}.
The 1:1 composition of LaAs was confirmed by energy dispersive x-ray spectroscopy using a JOEL field emission electron microscope quipped with an EDAX detector.
Powder x-ray diffraction (PXRD) was performed using a Bruker D8 ECO instrument.
FullProf suite was used for the Rietveld refinement of the PXRD data \cite{rodriguez-carvajal_recent_1993}.
Resistivity and the Hall effect were measured with a standard four probe technique in a Quantum Design Dynacool in both positive and negative field directions.
The data were symmetrized for transverse magnetoresistance (MR) and anti-symmetrized for the Hall effect.
Density functional theory (DFT) calculations with full-potential linearized augmented plane-wave (LAPW) method were implemented in the WIEN2k code \cite{blaha_wien2k_2001} with the basis-size controlling parameter $\mathrm{RK_{max}}=8.5$ and 10000 k-points.
Both the Perdew-Burke-Ernzerhof (PBE) \cite{perdew_generalized_1996} and the modified Becke-Johnson (mBJ) exchange-correlation potentials \cite{tran_accurate_2009} were used in the calculations with spin-orbit coupling (SOC).
ARPES measurements were performed at the high resolution branch of the i-05 beamline at Diamond Light Source.
Single crystals of LaAs were cleaved in an ultrahigh vacuum environment of $10^{-10}$ torr and measured at both 7 K and 220 K.
A Scienta R4000 electron analyzer was used with total energy and angular resolutions of 10 meV and $0.3^\circ$.



\section{\label{Results}Results}
\subsection{\label{compare} Magnetoresistance and Hall effect}

\begin{figure*}
\includegraphics[width=1\textwidth]{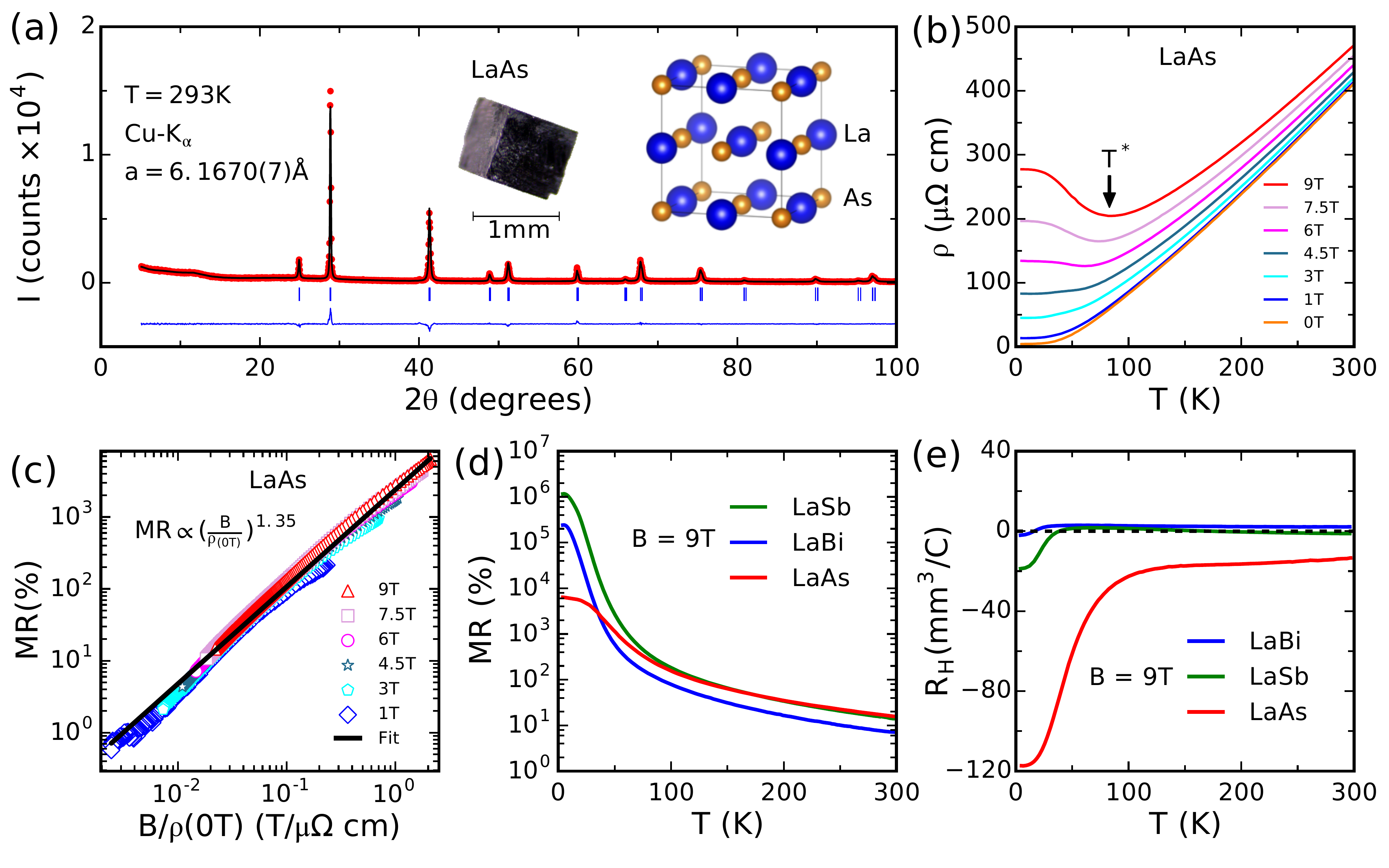}
\caption{\label{RT}
(a) Rietveld refinement on the powder x-ray data from LaAs in the space group $Fm\bar{3}m$ with $R_{wp}=7.71$, $R_{exp}=6.05$, and $\chi^2 =1.63$.
Inset shows a picture of the single crystal and a drawing of the LaAs unit cell.
(b) Resistivity as a function of temperature at different fields in LaAs.
(c) Kohler scaling analysis on the resistivity data.
(d) Magnetoresistance as a function of temperature in LaAs, LaSb, and LaBi on a logarithmic scale.
(e) Hall coefficient as a function of temperature in LaAs, LaSb, and LaBi.
}
\end{figure*}

Fig.~\ref{RT}(a) shows the face-centered cubic (fcc) structure of LaAs.
The high quality of crystals is confirmed by the absence of impurity phases in the x-ray pattern and the low $\chi^2$ in the Rietveld refinement.
Fig.~\ref{RT}(b) shows $\rho(T)$ in LaAs measured at different magnetic fields.
At $B=9$~T, with decreasing temperature, $\rho(T)$ decreases initially, then shows a minimum followed by an upturn, and eventually plateaus.
With decreasing magnetic field, the resistivity upturn gradually disappears.
Such behavior is a generic XMR profile \cite{ali_large_2014,tafti_temperaturefield_2016}.
The resistivity minimum at fields above 5 T in Fig.~\ref{RT}(b) can be understood by comparing the energy scale of cyclotron frequency $\hbar\omega_c=\hbar e B/m^*$ to the thermal energy $k_BT$.
As shown later, from quantum oscillations, the average effective mass on the small Fermi surfaces of LaAs is $m^*\approx 0.15\, m_e$.
Therefore, MR appears below $T^*=\hbar e B/m^*k_B \approx 80$ K (at $B=9$ T).
If cyclotron motion is the main source of resistivity upturn, MR at all temperatures and fields must follow the Kohler's scaling rule:
\begin{equation}
\label{kohler}
\mathrm{MR}(\%) = \frac{\rho(T,B)-\rho(T,0)} {\rho(T,0)} \times 100 \propto \left(\frac{B}{\rho(T,0)}\right)^\nu
\end{equation}
Fig.~\ref{RT}(c) shows the Kohler's law is obeyed in LaAs, ruling out a field-induced metal-insulator transition or a temperature-induced Lifshitz transition \cite{spain_kohlers_1976,wang_origin_2015}.
The presence of an XMR profile in the absence of a Lifshitz transition in LaAs is similar to LaSb \cite{zeng_compensated_2016} and LaBi \cite{kumar_observation_2016}.
However, XMR is orders of magnitude smaller in LaAs compared to LaSb and LaBi as shown in Fig.~\ref{RT}(d).
It is shown in prior work \cite{tafti_temperaturefield_2016} that the XMR magnitude correlates with the residual resistivity ratio (RRR).
Fig.~S1 in the Supplemental Material \cite{suppmatt} compares a LaAs and a LaBi crystal with similar RRR where the XMR is an order of magnitude smaller in LaAs.
At $B=9$~T, the low temperature resistivity is smaller than the room temperature resistivity $\rho(2\textrm{K})<\rho(300\textrm{K})$ in LaAs whereas $\rho(2\textrm{K})>\rho(300\textrm{K})$ in LaBi/LaSb (see Fig.~S1).
Since the large magnitude of XMR in LaSb and LaBi is attributed to perfect compensation between electrons and holes \cite{sun_large_2016,zeng_compensated_2016}, we measured the Hall effect to examine the compensation in LaAs.
Fig.~\ref{RT}(e) shows the Hall coefficient (\RH) in LaAs acquires a much larger negative magnitude without sign change, different from LaSb and LaBi.
At first glance, this may suggest that LaAs is not compensated.
However, our detailed analyses below show that LaAs is as compensated as LaSb/LaBi, and the difference in \RH~comes from an order of magnitude difference in the relative mobilities of electrons and holes (\emph{mobility mismatch}) instead of their concentrations.
Next, we turn to ARPES to map the Fermi surfaces of LaAs and to investigate signatures of topological band structure.


\subsection{\label{arp} ARPES}
\begin{figure*}[t]

\includegraphics[width=1\textwidth,center]{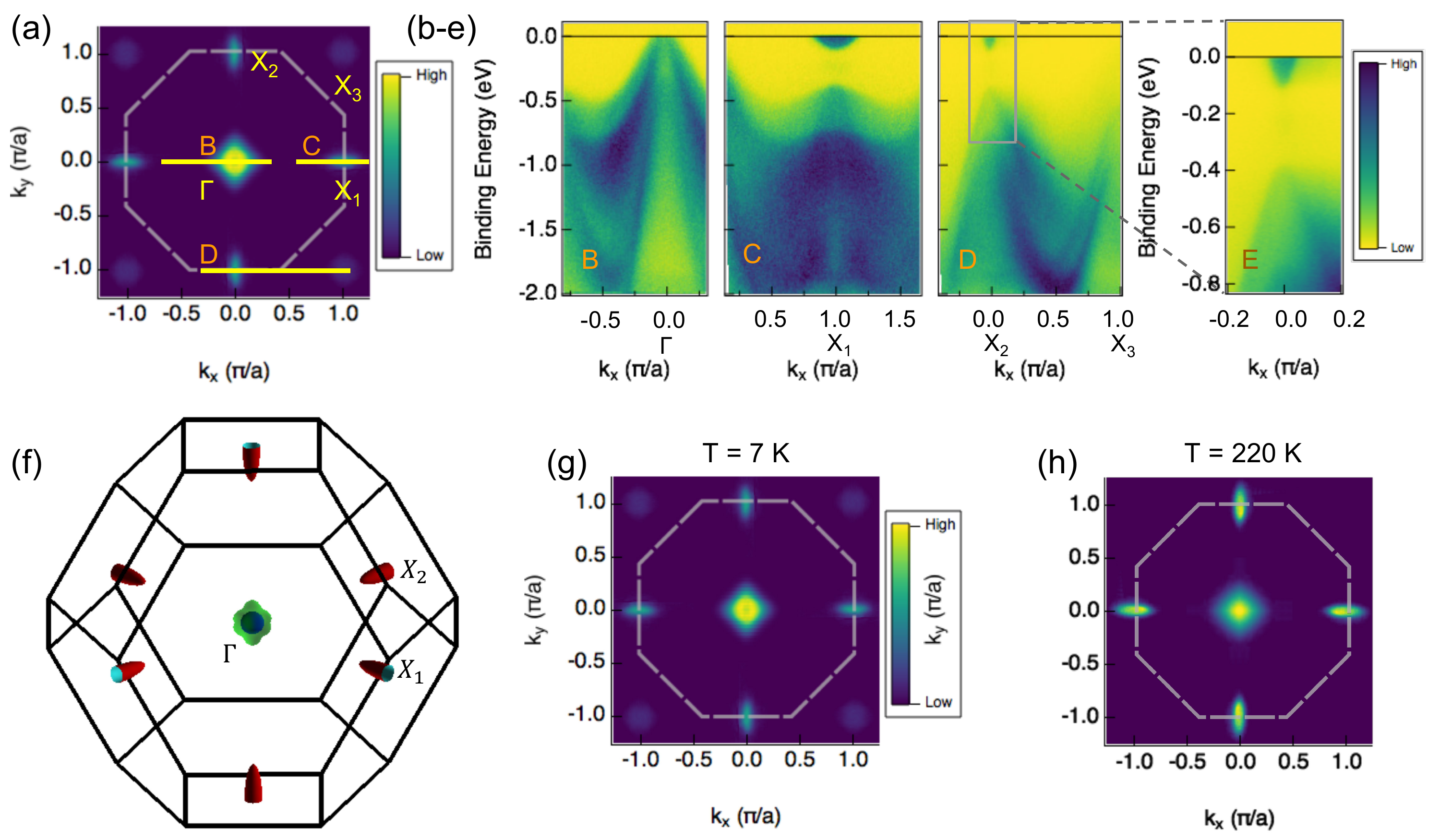}
\caption{\label{ARPES}
(a) Constant energy surfaces of LaAs taken at $E_F$ in the $k_x-k_y$ plane.
The dashed overlay is the first Brillouin zone, the solid lines indicate the respective locations of the dispersion cuts in (B),(C), and (D).
Cross-sections of the ellipsoidal electron pockets ($\alpha$) are visible at $X_1$, $X_2$, and $X_3$.
Cross-sections of the the hole pockets ($\beta$ and $\gamma$) are visible at $\Gamma$.
(b) Dispersion along $\Gamma-X_1$, centered on the hole bands.
(c) Dispersion along $\Gamma - X_1$ centered on the electron pocket along the major axis of the ellipsoid.
There is no band crossing along this direction.
(d) Momentum dispersive cuts along the minor axis of the ellipsoid electron pocket ($X_2-X_3$ direction).
(e) Zoomed-in dispersion along $X_2-X_3$ at $T = 7$ K conforming the absence of a Dirac cone.
(f) The Fermi surface of LaAs from DFT calculations in agreement with the ARPES picture.
(g) Symmetrized constant energy surfaces taken at $E_F$ of LaAs at $T = 7$ K.
(h) The same view at $T = 220$ K.
}
\end{figure*}

%
Prior ARPES studies suggest a progression from topological to non-topological band structure in lanthanum monopnictides with decreasing pnictogen size.
LaBi has topological band inversion with Dirac cones \cite{nayak_multiple_2017,lou_evidence_2017,niu_presence_2016}.
LaSb appears to be on the verge of a transition from topological to trivial band structure \cite{niu_presence_2016,zeng_compensated_2016,oinuma_three-dimensional_2017}.
Here, we investigate the case for LaAs.
Fig.~\ref{ARPES}(a) is a 2D constant energy surface at $E_F$, symmetrized to fill the entire Brillouin zone.
LaAs has ellipsoidal pockets at the faces of the fcc Brillouin zone ($X$ points) and two concentric spheroidal pockets at the center of the zone ($\Gamma$ point), similar to the Fermi surfaces of LaSb and LaBi \cite{zeng_compensated_2016,nayak_multiple_2017}.
Figs.~\ref{ARPES}(b-e) show the measured dispersions along three paths (B), (C), and (D) as indicated on Fig.~\ref{ARPES}(a).
Path (B) is along $\Gamma-X_1$, centered around $\Gamma$, showing that the two concentric pockets at $\Gamma$ come from two hole bands.
Path (C) is also along $\Gamma-X_1$, but centered around $X_1$, showing the major axis of the ellipsoidal pocket which clearly comes from an electron band.
Path (D) is along $X_2-X_3$, showing the minor axis of the ellipsoidal electron pocket.
The clear lack of a band crossing in all cuts precludes the existence of topological states in LaAs.
Fig.~\ref{ARPES}(e) zooms in the dispersion along the $X_2-X_3$ direction to highlight the clear gap beneath the $\alpha$ pocket with no evidence for a band crossing or a Dirac cone.
These results demonstrate a transition in the lanthanum monopnictide family, from LaBi with topological band structure where Dirac cones are present, to LaAs with trivial band structure where Dirac cones are absent.
In the Supplemental Material \cite{suppmatt}, we present the dispersion of the electronic states at $X$ along the sample normal direction ($k_z$) to confirm their periodicity and the absence of surface states.
Fig.~\ref{ARPES}(f) renders the three dimensional Brillouin zone of LaAs with the electron pockets ($\alpha$) at $X$ and the  hole pockets (inner $\beta$ and outer $\gamma$) at $\Gamma$ from DFT calculations.
Figs.~\ref{ARPES}(g) and (h) show a comparison of the Fermi surfaces measured in LaAs at $T=9$ K and $T=220$ K.
The largely unchanged Fermi surfaces observed by ARPES rule out a Lifshitz transition in LaAs as a function of temperature, consistent with the Kohler scaling of the resistivity data in Fig.~\ref{RT}(c).
Next, we discuss the DFT calculations that lead to Fig.~\ref{ARPES}(f).
%


\subsection{\label{bands} Band Structure}
\begin{figure}
\includegraphics[width=3.5in]{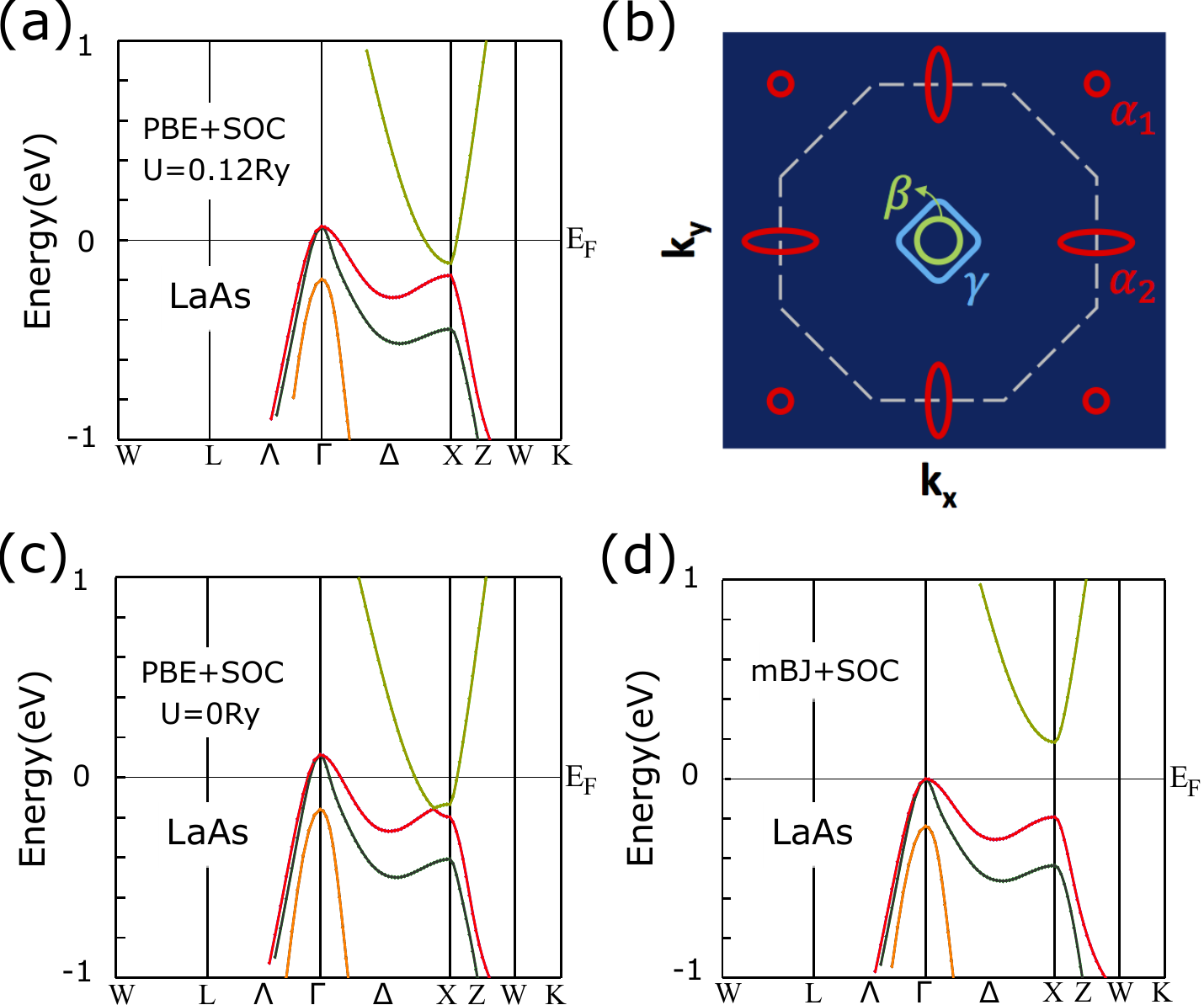}
\caption{\label{BS}
(a) The correct band structure of LaAs (consistent with ARPES and quantum oscillations) from a PBE+SOC+U calculation with $U=0.12$ Ry.
(b) Schematic 2D map of the Brillouin zone with ellipsoidal electron pockets ($\alpha$) at the $X$ points and concentric hole pockets (inner $\beta$ and outer $\gamma$) at the $\Gamma$ point.
(c) Band structure of LaAs calculated by PBE+SOC showing a band crossing near $X$.
(d) Band structure of LaAs calculated by mBJ+SOC showing a large gap that lifts the electron pocket from $E_F$.
}
\end{figure}

As presented in the previous section, ARPES measurements along $\Gamma-X$, shown in Fig.~\ref{ARPES}, revealed two hole bands at $\Gamma$ and one small electron pocket without band crossing at $X$.
To capture these features, we performed a PBE+SOC+U calculation to open a gap at $X$ while maintaining the position of the electron band bottom below $E_F$ as shown in Fig.~\ref{BS}(a).
The sizes of the gap and the electron pocket are tuned by varying $U$.
Our choice of $U = 0.12$ Ry is justified by the size of the Fermi pockets determined by quantum oscillations as described in the next section.
Fig.~\ref{BS}(b) is a schematic illustration of the fcc Brillouin zone of LaAs in the $k_x-k_y$ plane.
The larger ($\alpha_1$) and the smaller ($\alpha_2$) cross sections of the electron pockets appear at the $X$ points.
The smaller ($\beta$) and the larger ($\gamma$) hole pockets appear at the $\Gamma$ point.
Due to the small sizes of LaAs Fermi surfaces, DFT calculations could easily produce misleading results.
For example, Fig.~\ref{BS}(c) shows the outcome of a PBE+SOC calculation on LaAs.
This calculation correctly captures the band structure of LaBi \cite{niu_presence_2016,kumar_observation_2016}.
However, in LaAs, it overestimates the $\alpha$ pocket size and incorrectly predicts a band crossing at $X$.
Fig.~\ref{BS}(d) shows the outcome of a mBJ+SOC calculation on LaAs.
This calculation accurately describes LaSb according to ARPES and transport experiments \cite{tafti_resistivity_2016,zeng_compensated_2016}.
However, in LaAs, it predicts that the electron pocket at $X$ is lifted from the Fermi level, contradicting both the existence of the electron $\alpha$ pockets in Fig.~\ref{ARPES}(a) and the observed negative Hall effect in Fig. \ref{RT}(e).
Despite the simple rock-salt structure of lanthanum monopnictides, it is challenging to correctly predict their band structures without experimental guidance.
Indeed, a prior theoretical DFT study incorrectly predicted LaAs to be a semicondctor with 0.1 eV gap \cite{yan_theoretical_2014}.


\subsection{\label{oscillations} Quantum Oscillations}
\begin{figure*}
\includegraphics[width=1\textwidth,center]{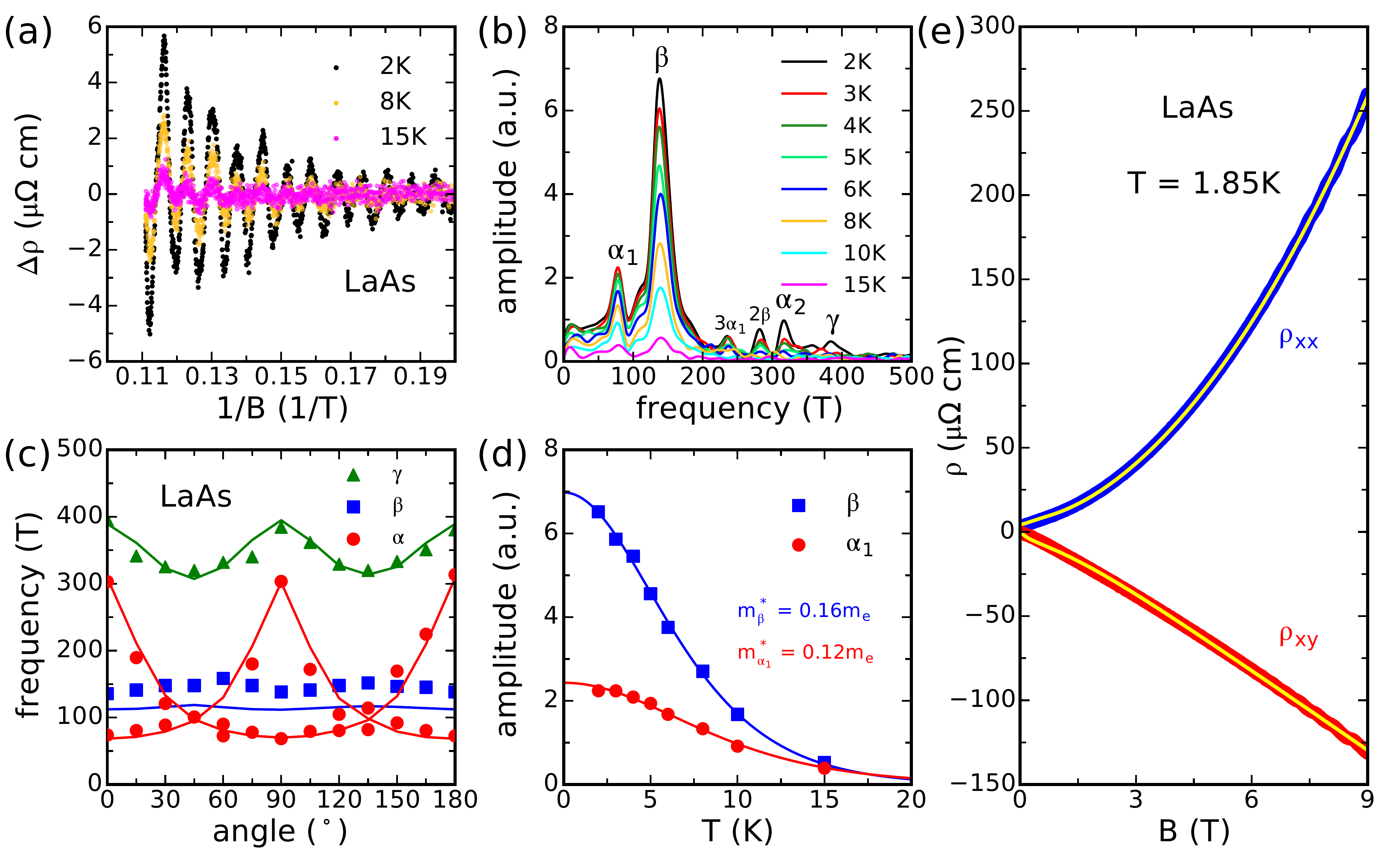}
\caption{\label{QO}
(a) The oscillatory part of resistivity $\Delta \rho$ plotted as a function of $1/B$ at three representative temperatures.
(b) Fast Fourier Transform (FFT) of $\Delta \rho$ data at different temperatures.
Four main frequencies ($\alpha_1$, $\alpha_2$, $\beta$, $\gamma$) and their harmonics ($3\alpha_1$, $2\beta$) are identified.
(c) The observed angular dependence of the main frequencies (solid symbols) agrees with the calculated results from DFT (solid lines).
(d) Lifshitz-Kosevich fit to the temperature dependence of the FFT amplitudes.
The effective masses of $\alpha_1$ and $\beta$ are extracted reliably within the resolution of our data.
(e) Multiband fit implemented simultaneously to the resistivity ($\rho_{xx}$) and the Hall effect ($\rho_{xy}$) as a function of field.
}
\end{figure*}

For a precise measurement of the sizes of electron and hole pockets in LaAs, we studied quantum oscillations in the resistivity channel known as the Shubnikov-de Haas effect.
Due to the small size of the Fermi surfaces in semimetals such as LaAs, it is challenging to reliably extract the Fermi volumes from the ARPES spectra as can be seen in Fig.~\ref{ARPES}.
For example, the electron to hole carrier concentration in YSb is estimated to be $n_e/n_h=0.81$ (moderate compensation) from ARPES~\cite{he_distinct_2016} whereas $n_e/n_h=0.95$ (almost perfect compensation) from quantum oscillations~\cite{xu_origin_2017}.
Fig.~\ref{QO}(a) shows the oscillatory part of resistivity $\Delta \rho$ after removing a smooth background from the resistivity data at different temperatures.
Oscillations are periodic in $1/B$ and their amplitudes decrease with increasing temperature.
Fig.~\ref{QO}(b) shows the Fast Fourier Transform (FFT) spectrum of the oscillations at different temperatures.
FFT peaks at $\alpha_1=76$ T and $\alpha_2=315$ T correspond to the smaller and the larger areas of the ellipsoidal electron pocket ($\alpha$).
The peaks at $\beta=140$ T and $\gamma=382.5$ T correspond to the smaller ($\beta$) and the larger ($\gamma$) hole pockets.
These frequencies were used to tune the $U$ in PBE+SOC+U calculation (Fig.~\ref{BS}(a)) until the calculated frequencies from DFT matched the experimental frequencies (Supplemental Material \cite{suppmatt}).
Angular dependence of the FFT peaks is used to assign the frequencies to $\alpha$, $\beta$, and $\gamma$ pockets.
Fig.~\ref{QO}(c) shows a strong angle dependence for the $\alpha$ frequencies as expected from the minor ($\alpha_1$) and the major ($\alpha_2$) extremal areas of the ellipsoidal pocket \cite{rourke_numerical_2012}.
The $\beta$ frequency is angle independent as expected from a spherical pocket \cite{rourke_numerical_2012}.
The $\gamma$ frequency with a mild angle dependence corresponds to a jack-shaped pocket as illustrated in Fig.~\ref{ARPES}(f).
Solid lines on Fig.~\ref{QO}(c) represent calculated frequencies for LaAs from DFT using the SKEAF program \cite{rourke_numerical_2012}.
The agreement between calculated and observed frequencies at different angles confirms the Fermi surface geometry.
Using the Onsager relation $F=\frac{\phi_0}{2\pi^2}A_{ext}$, where $\phi_0$ is the quantum of flux, we extracted the extremal orbit areas $A_{ext}$ for $\alpha$, $\beta$, and $\gamma$, then calculated their volumes to find the number of carriers in each pocket (Supplemental Material \cite{suppmatt}).
As a result, $n_{\alpha}=1.55 \times 10^{19}$, $n_{\beta}=0.94 \times 10^{19}$, and $n_{\gamma}=3.66 \times 10^{19}$ cm$^{-3}$, corresponding to $n_e/n_h=1.01$.
A similar analysis on LaSb yields $n_e/n_h=0.99$ \cite{zeng_compensated_2016}.
Therefore, LaAs is as compensated as LaSb.
The effective masses of the carriers on $\alpha_1$ and $\beta$ surfaces are estimated by fitting the FFT amplitudes to the Lifshitz-Kosevich formula \cite{shoenberg_magnetic_2009,willardson1967physics} in Fig.~\ref{QO}(d).
The average mass, $m^*\approx 0.15\,m_e$, used earlier to estimate $T^*$ in Fig.~\ref{RT}(b), came from this analysis.


\subsection{\label{discussion} Discussion}

The most striking difference between LaAs and the other members of its chemical family, LaSb and LaBi, is the significant reduction in the XMR magnitude of LaAs (Fig.~\ref{RT}(d)).
Our goal is to understand this dramatic reduction of XMR magnitude in LaAs through the lens of the various probes presented thus far.
The Hall effect data in Fig.~\ref{RT}(e) showed that $R_H (T)$ had a much larger amplitude in LaAs with no change of sign, different from LaSb/LaBi.
This could suggest a lack of compensation in LaAs, a proposed prerequisite for XMR \cite{zeng_compensated_2016}.
However, ARPES (Fig.~\ref{ARPES}) qualitatively showed comparable electron and hole pockets, and quantum oscillations (Fig.~\ref{QO}) quantitatively confirmed their compensated densities in LaAs similar to LaSb/LaBi \cite{zeng_compensated_2016,Hasegawa_fermi_1985,sun_large_2016}.
To further investigate this, we implemented a multiband fit to the field dependence of $\rho_{xx}$ and $\rho_{xy}$ simultaneously, as shown in Fig.~\ref{QO}(e) and elaborated in Supplemental Material \cite{suppmatt}.
Our model assumed three electron pockets and two hole pockets, analogous to LaSb/LaBi, and supported by both our ARPES and quantum oscillations measurements.
This multiband fit predicted $n_e/n_h=1.005$ in LaAs, strengthening the consensus around compensation.
To explain the large discrepancies between $R_H(T)$ in the three compounds, we appeal to the mobility mismatch between electron and hole carriers.
From the multiband fits in Fig.~\ref{QO}(e), the average electron to hole mobility ratio $\mu_e/\mu_h \approx 13$ in LaAs.
This is an order of magnitude different from $\mu_e/\mu_h \approx 1$ in LaSb/LaBi \cite{zeng_compensated_2016,kumar_observation_2016}.
For a more intuitive understanding of the impact of such mobility mismatch on $R_H$, we turn to the two-band model expression for the Hall resistivity \cite{ashcroft_solid_1976}:
\begin{align}
\label{twob}
\rho_{xy}=\frac{(R_h\rho_e^2+R_e\rho_h^2)B+(R_hR_e^2+R_eR_h^2)B^3}{(\rho_h+\rho_e)^2+(R_h+R_e)^2B^2}
\end{align}
where $R_{h(e)}$ and $\rho_{h(e)}$ stand for the Hall coefficient and the resistivity of an isolated hole (electron) band.
In the limit of compensation, where $n_e/n_h=1$, Eq.~\ref{twob} reduces to a simple form for the Hall coefficient ($R_H=\rho_{xy}/B$):
\begin{equation}
\label{mus}
R_H=\frac{1}{ne}\frac{\mu_h-\mu_e}{\mu_h+\mu_e}
\end{equation}
From here, we attribute the larger magnitude of $R_H$ in LaAs (Fig.~\ref{RT}(e)) to the smaller Fermi surfaces i.e. smaller $n$, and we attribute the lack of sign change in LaAs to the mobility mismatch i.e. $\mu_e \neq \mu_h$.
LaAs, LaSb, and LaBi are all nearly compensated semimetals which exhibit XMR, albeit to varying magnitudes.
Therefore, electron-hole compensation cannot be the cause for the significant reduction of XMR magnitude in LaAs when compared to its siblings.
We argue instead that one key quantity for determining XMR magnitude in these compensated materials is the matching of electron and hole mobilities.
A mobility mismatch allows for a larger Hall field to develop under applied magnetic fields.
This larger Hall field in LaAs counteracts the Lorentz force more effectively and disrupts the field induced cyclotron motion, therefore reduces the XMR magnitude.


\subsection{\label{conclusions} Conclusions}

By growing and characterizing single crystals of LaAs, we confirmed the qualitative existence of XMR in this material although the magnitude is quantitatively much reduced.
Quantum oscillations, multiband fit, and ARPES measurements confirm that LaAs is almost perfectly compensated, similar to LaSb/LaBi.
The multiband fit shows that the larger Hall field and the smaller MR in LaAs are due to the electron-hole mobility mismatch instead of a lack of compensation.
The challenges of band structure calculations for semimetals with small Fermi surfaces are highlighted by presenting three different DFT calculations on LaAs with three different results.
The correct calculation comes from a PBE+SOC+U scheme by tuning $U$ until the calculated Fermi surfaces match the experimental observations.
The ARPES measurements resolve a non-topological band structure in LaAs, placing it on the other side of a topological transition from LaBi.
This is the first presentation of a transition from topological to non-topological band structure in the lanthanum monopnictide family.
The existence of the XMR resistivity profile in all three materials must therefore result from compensation and independent of topology.
Alternative explanations for XMR such as a field induced metal-insulator transition are also ruled out by confirming the Kohler scaling on the resistivity data and by showing nearly identical ARPES maps at $T=7$ and 220~K.


\section*{ACKNOWLEDGMENTS}

F.T. is grateful to P.~Rourke from NRC for discussions on SKEAF.
T.N. would like to thank the NSF Graduate Research Fellowship Program for support during this work.
The work at BC is funded by the NSF Grant No. DMR-5104811. The work at CU is funded by the DOE Office of Basic Energy Sciences under grant DE-FG02-03ER46066. The Diamond light source is funded as a joint venture by the UK Government through the Science and Technology Facilities Council (STFC) in partnership with the Wellcome Trust.





\bibliography{Yang_LaAs_01dec2017}

\end{document}